\documentclass[12pt,a4paper]{article}

\usepackage{amssymb}
\usepackage{amsmath, epsfig, color, subfigure,bm}
\usepackage{fullpage}
\usepackage{color}
\usepackage{epsfig}
\usepackage{epstopdf}
\usepackage{subfigure}
\usepackage{natbib}

\newcommand{\be}{\begin{equation}}
\newcommand{\en}{\end{equation}}

\renewcommand{\vec}[1]{\boldsymbol{#1}}

\begin{document}

\title{Deficiencies in numerical models\\ of anisotropic nonlinearly elastic materials}

\author{A. N\'{i} Annaidh$^a$, M. Destrade$^{a,b}$, M.D. Gilchrist$^a$, J.G. Murphy$^{c,b,\star}$. \\[8pt]
$^a$School of Mechanical \& Materials Engineering, \\
University College Dublin, Belfield, Dublin 4, Ireland; \\[4pt]
$^b$School of Mathematics, Statistics and Applied Mathematics, \\
National University of Ireland Galway, Ireland; \\[4pt]
$^c$Centre for Medical Engineering Research, \\
Dublin City University, Glasnevin, Dublin 9, Ireland.}

\date{$^\star$ Corresponding author. \\ email: jeremiah.murphy$@$dcu.ie, phone: +353-1-700-8924}

\maketitle

\begin{abstract}
 
Incompressible nonlinearly hyperelastic materials are rarely simulated in Finite Element numerical experiments as being perfectly incompressible because of the numerical difficulties associated with globally satisfying this constraint. Most commercial Finite Element packages therefore assume that the material is slightly compressible. It is then further assumed that the corresponding strain-energy function can be decomposed additively into volumetric and deviatoric parts.  
We show that this decomposition is not physically realistic, especially for anisotropic materials, which are of particular interest for simulating the mechanical response of biological soft tissue.  
The most striking illustration of the shortcoming is that with this decomposition, an anisotropic cube under hydrostatic tension deforms into another cube instead of a hexahedron with non-parallel faces.
Furthermore, commercial numerical codes require the specification of a `compressibility parameter' (or `penalty factor'), which arises naturally from the flawed additive decomposition of the strain-energy function. 
This parameter is often linked to a `bulk modulus', although this notion makes no sense for anisotropic solids; we  show that it is essentially an arbitrary parameter and that infinitesimal changes to it result in significant changes in the predicted stress response. This is illustrated with numerical simulations for biaxial tension experiments of arteries, where the magnitude of the stress response is found to change by several orders of magnitude when infinitesimal changes in `Poisson's ratio' close to the perfect incompressibility limit of $1/2$ are made.

\end{abstract}

\newpage

%%%%%%%%%%%%%%%%%%%%%%%%%%%

\section{Introduction}

%%%%%%%%%%%%%%%%%%%%%%%%%%%

Since biological systems inevitably involve complex geometries, microstructure and boundary conditions, numerical simulations of the mechanical response of soft tissue are required in order to estimate the stress distribution. The Finite Element Method is typically the preferred numerical method and, because of the complexity, commercial (such as ABAQUS, ANSYS, LS-DYNA, MSC.Nastran, etc.) or non-commercial (such as FEBio, CalculiX, etc.) formulations of this method are usually employed to predict the stress distribution for soft tissue. It is worth noting that \citet{Eea} report an almost exponential increase over the last 40 years in published studies utilizing finite element analysis as a research tool. This increased reliance on using computational models for analysis has been achieved mainly by rapid advances in non-invasive medical imaging such as Computational Tomography and Magnetic Resonance Imaging being coupled with the use of computer software for automatically meshing complex anatomical structures and for solving large scale problems quickly. However, these rapid technological advances in imaging and computer engineering have not been matched by similarly rapid advances in experimental methods to characterise soft matter. Indeed, there is a dearth of accurate mechanical property data for soft biological tissue such as skin, muscle and neural tissue and sufficiently detailed data is only recently beginning to be obtained \citep{N1, N2, Rea}. As more sophisticated medical devices continue to be designed for implantation into soft tissue, and as computer assisted/guided surgery relies on increasingly sophisticated models for controlling the mechanical interaction between surgical instruments and soft tissue, the need to understand the behaviour of incompressible nonlinearly hyperelastic materials is becoming increasingly urgent.

The deformations of biological, soft tissue are usually \emph{assumed} to be accompanied by only infinitesimal volume changes due to the tissue's high water content \citep{VaD}. 
However, there is no generally agreed method for modelling slightly compressible materials. The most popular approach, and that implemented in most commercial Finite Element codes, is to decompose the strain-energy function additively into volumetric and deviatoric parts, with the deviatoric part constructed in such a way that only isochoric deformations can be considered. A summary of this approach can be found, for example, in  \citet{Book} and also in Section \ref{Preliminaries} here.
This decomposition has the advantage of having an innate intuitive appeal and, as importantly, is mathematically convenient. However, the physical basis for this assumption has rarely been tested. Exceptions include the work of \cite{San} and \cite{HaM} who proved that for \emph{isotropic} materials this decomposition is equivalent to assuming that the hydrostatic Cauchy stress is a function only of the invariant measuring volume change for every deformation (see Section \ref{Isotropic materials}). 
Following \cite{San}, who proved the same result for orthotropic materials,  it is shown in Section \ref{add} that this result also holds for non-linearly hyperelastic \emph{anisotropic} materials with two preferred directions, the standard phenomenological model for large, elastic arteries.  
The consequences of this identity are then explored, with the conclusion being that the additive decomposition of strain-energy functions into volumetric and deviatoric parts in order to model slightly compressible materials is valid only for isotropic materials under hydrostatic tension/compression and is \emph{not} appropriate for anisotropic materials. 
The limited experimental data available, due to  \cite{Penn}, support this view.  
Therefore Finite Element analyses that use the volumetric and deviatoric decomposition as a fundamental constitutive assumption are not based on good physics for anisotropic soft tissues and, consequently, the corresponding stress predictions must be viewed skeptically. 

Even if one were to ignore our concerns about this decomposition and continue to use commercial Finite Element codes based on it, there is a serious practical problem associated with using these codes. 
Most codes assume that the hydrostatic stress is proportional to a function of the assumed infinitesimal volume change but there is no guidance as to how the constant of proportionality should be chosen; in effect, it is an arbitrary parameter. 
Most commercial codes do provide a default value. For example, ABAQUS \citep{AUM} assumes, without justification, the value $\kappa/\mu = 20$ for its artery model, where $\kappa$ is defined to be the bulk modulus, remembering that we dispute that this is well-defined for anisotropic materials, and $\mu$ is defined to be the largest value of the initial shear moduli among the different material directions. 
This value seems to have been directly imported from the isotropic formulation, where $\kappa/\mu = 20$ corresponds to a Poisson's ratio $\nu$ of 0.475. 
Repeating this implicit identification of the anisotropic compressibility factor with its isotropic counterpart, we show here that for biaxial experiments on a nonlinear, homogenous material with two preferred directions introduced by \cite{Hea} and by  \cite{Gea}, there are significant variations in the predicted stress response as a result of variations in this arbitrary parameter induced by infinitesimal variations in Poisson's ratio, \emph{even for small strains} (see Section \ref{Finite Element simulations}). 
This is worrying for two reasons: first, there is the issue of reproducibility of numerical experiments, since many reported simulations do not reveal the value of the compressibility constant used. 
The more important and related second problem is that Finite Element predictions of stress based on additive decomposition must now be viewed even more skeptically in terms of their correspondence to reality because the experimental determination of this crucial parameter is hardly ever conducted.
Thus, at the very best, only a qualitative estimate of the physical stress can be obtained and even this cannot be stated with much confidence, because the principal stresses for slightly compressible models are hyper-sensitive to variations in Poisson's ratio, see Section \ref{sen}.
This has potentially serious implications, in particular, for the numerical modelling of biological, soft tissue where Finite Element analysis is a basic tool.

Companies that sell Finite Element codes seem complacent when it comes to the issue of incompressibility. For example, the LS-DYNA manual states that for their MAT\_SOFT\_TISSUE model, used to represent transversely isotropic biological soft tissue,
\begin{quotation}
\noindent 
the bulk modulus $K$ should be at least 3 orders of magnitude larger than $C_1$ [shear modulus] to ensure near-incompressible material behavior.
\end{quotation}
This would correspond to a Poisson's ratio between 0.4995 and 0.5.
However, for the MAT\_BLATZ-KO\_RUBBER model, the manual states
\begin{quotation}
\noindent
this one parameter material allows the modeling of nearly incompressible continuum rubber. The Poisson's ratio is fixed to 0.463.
\end{quotation}
Thus the bulk modulus here is only 13 times larger than the shear modulus.
Similarly, the ABAQUS manual states that
 \begin{quotation}
\noindent in applications where the material is not highly confined, the degree of compressibility is typically not crucial.
\end{quotation}
We show here, however,  that the degree of compressibility is indeed very crucial, even for the  unconfined problem of biaxial tension. The crucial dependence on the compressibility parameter in commercial Finite Element codes has been demonstrated previously by \cite{Gent} and by \cite{Dea}, who showed that in shearing deformations of isotropic materials the normal stress distribution can exhibit extreme sensitivity to changes in Poisson's ratio. The seemingly obvious solution to this dependence on a compressibility factor of simply simulating perfect incompressibility is not a valid approach since no material is perfectly incompressible. The analysis presented here suggests that the difference between predicted stress distributions assuming this idealisation and those obtained assuming slight compressibility is likely to be very significant. The solution to the problems identified here must begin with extensive and careful experimentation of anisotropic materials, and, in particular, biological, soft tissue, to determine the variation of compressibility under mechanical loading. It is only then that a rational constitutive framework that models this compressibility for anisotropic materials can be formulated.

\begin{figure}[ht]
\centering
\subfigure[Linear elastic implementation]{\includegraphics[trim = 10mm 0mm 100mm 5mm, clip, scale=0.39]{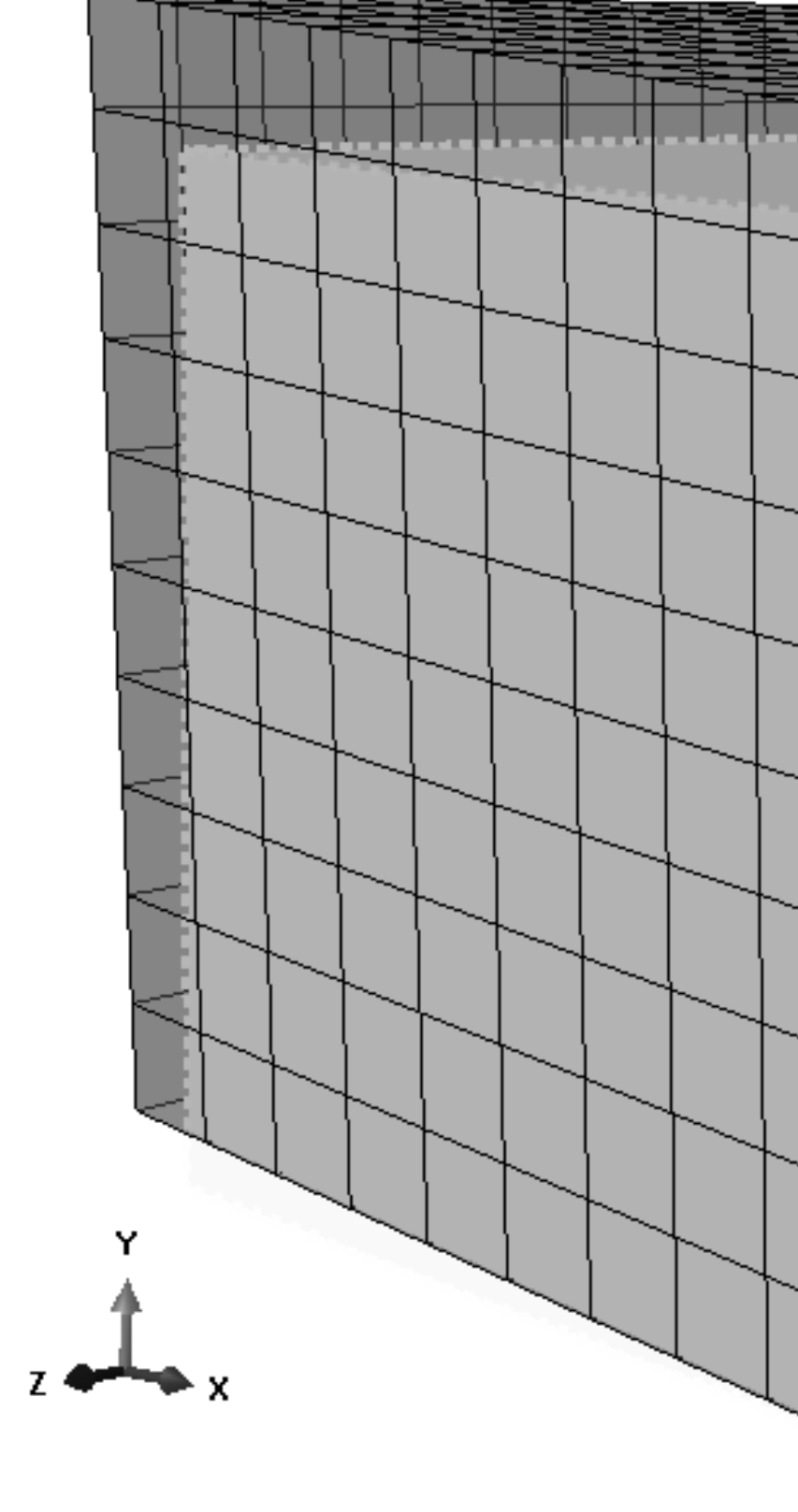} }
\subfigure[Hyperelastic implementation]{\includegraphics[trim = 10mm 0mm 100mm 5mm, clip, scale=0.38]{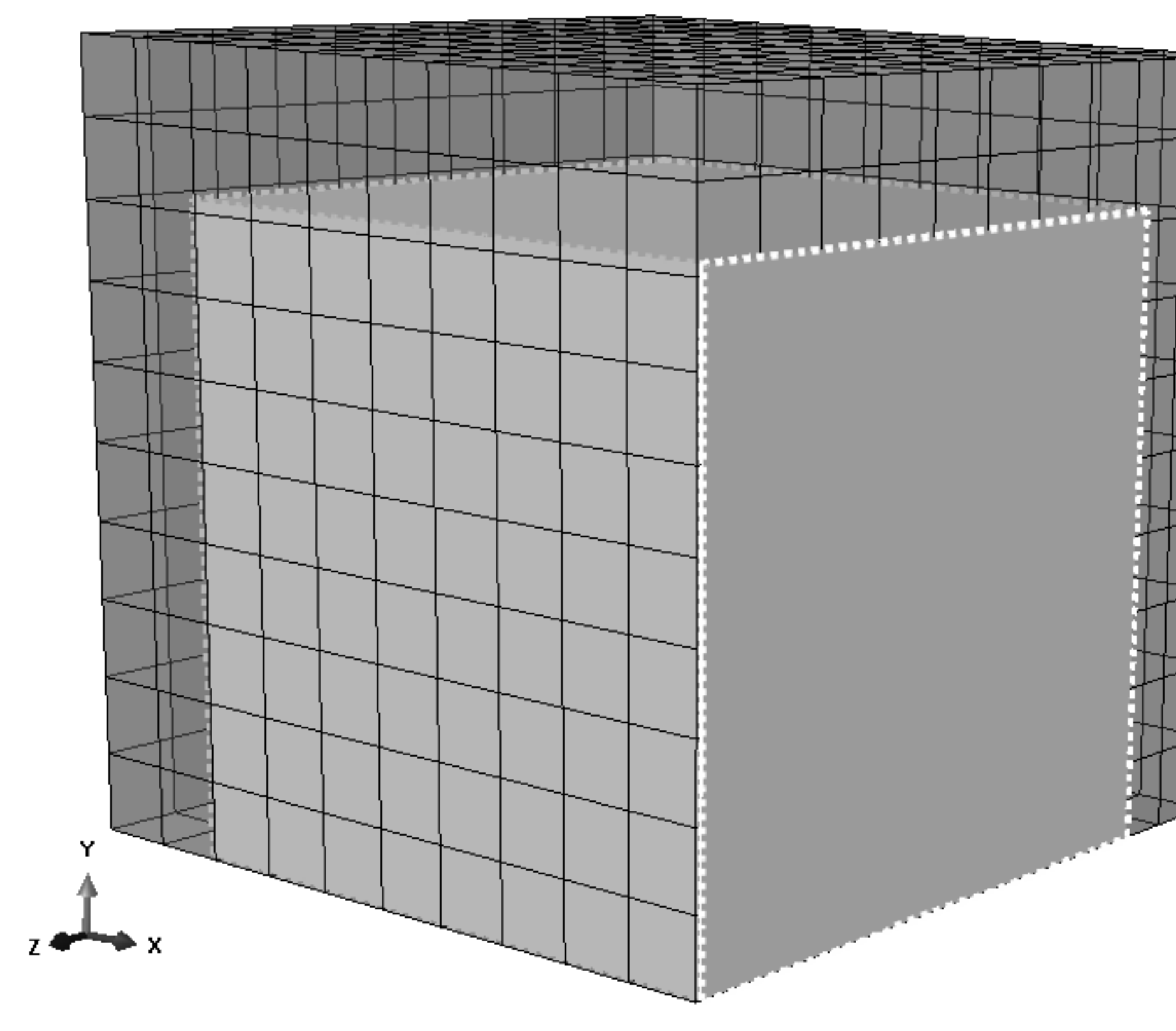} }
\caption{Deformed cube under hydrostatic tension. (a) Using the linear elastic implementation and the elastic constants of Zinc (orthotropic), the cube deforms into a hexahedron with non-parallel faces. (b) Using the hyperelastic implementation and material constants of arteries, the cube deforms into another larger cube.}
\label{fig:cube}
\end{figure}

We conclude this section with a simple, yet telling, experiment in ABAQUS, where we subject a cube of side 10 mm, meshed with 1000 C3D8R elements, to hydrostatic tension.
For our first experiment, we use the code to model an anisotropic solid which is characterized by `\emph{linear elastic behavior}': we take a cube made of orthotropic Zinc, with the following elastic constants $D_{1111} = 165$, $D_{1122}=31.1$, $D_{2222}=165$, $D_{1133}=50$, $D_{2233}=50$, $D_{3333}=61.8$, $D_{1212}=66.95$, $D_{1313}=D_{2323}=39.6$ (GPa) \citep{Hearmon79}. We rotate its symmetry axes by 45$^\circ$ with respect to its edges. We subject it to a hydrostatic tension of magnitude 10 GPa. 
In the second experiment, ABAQUS is used to model a solid  with  `\emph{anisotropic hyperelastic behavior}': we use the material constants of Eq.\eqref{conval} (but with no fibre dispersion, i.e. $\kappa_0=0$), with a `bulk modulus' of 150 kPa (see first line of Table \ref{Table1}). We subject it to a hydrostatic tension of magnitude 100 kPa.
In the first case, the cube deforms into an hexahedron with non-parallel faces, as it should, whilst in the second case, the cube deforms into another cube, of side 12.31 mm.
Figure \ref{fig:cube} clearly highlights a major problem with the implementation of a rational model of nonlinear  anisotropic elasticity into ABAQUS. This is discussed further in Sections \ref{Isotropic materials} and \ref{add}.

%%%%%%%%%%%%%%%%%%%%%%%%%%%%%%

\section{Preliminaries}
\label{Preliminaries}

%%%%%%%%%%%%%%%%%%%%%%%%%%%%%%

The nominal and Cauchy stress tensors ($\vec{S},\vec{\sigma}$ respectively) are related by
$$
\vec{\sigma}=J^{-1}\vec{FS},
$$
where $J \equiv \lambda_1\lambda_2\lambda_3$, with $\lambda_i$ the principal stretches, and $\vec{F}$ is the deformation gradient tensor. 
 \cite{Cism} gives the following constitutive law for compressible, homogeneous, non-linearly elastic materials with two preferred directions along the unit vectors $\vec{M}$, $\vec{M}'$   in the undeformed configuration:
\begin{multline}
\vec{S}=2W_1\vec{F}^T+2W_2\left( I_1\vec{I}-\vec{C}\right)\vec{F}^T+2I_3W_3\vec{F}^{-1}+2W_4\vec{M} \otimes \vec{FM}  \\
 + 2W_5\left( \vec{M} \otimes \vec{FCM}+\vec{CM} \otimes \vec{FM} \right) + 2W_6\vec{M}^{\prime} \otimes \vec{FM}^{\prime}   \\
+ 2W_7\left( \vec{M}^{\prime} \otimes \vec{FCM}^{\prime}+\vec{CM}^{\prime} \otimes \vec{FM}^{\prime} \right) + W_8\left( \vec{M} \otimes \vec{FM}^{\prime}+\vec{M}^{\prime} \otimes \vec{FM} \right),
\end{multline}
so that
\begin{multline} \label{cs}
J\vec{\sigma}=2W_1\vec{B}+2W_2\left( I_1\vec{B}-\vec{B}^2\right)+2I_3W_3\vec{I}+2W_4\vec{FM} \otimes \vec{FM}  \\
+ 2W_5\left( \vec{FM} \otimes \vec{BFM}+\vec{BFM} \otimes \vec{FM} \right) + 2W_6\vec{FM}^{\prime} \otimes \vec{FM}^{\prime}  \\
+ 2W_7\left( \vec{FM}^{\prime} \otimes \vec{BFM}^{\prime}+\vec{BFM}^{\prime} \otimes \vec{FM}^{\prime} \right) + W_8\left( \vec{FM} \otimes \vec{FM}^{\prime}+\vec{FM}^{\prime} \otimes \vec{FM} \right).
\end{multline}
Here $W=W\left(I_1,I_2,I_3,\dots, I_8\right)$ is the strain-energy function per unit undeformed volume, attached subscripts denote partial differentiation with respect to the appropriate principal strain invariant or pseudo-invariant, and $\vec{B}, \vec{C}$ are the left and right Cauchy-Green strain tensors respectively. The invariants are defined by
\begin{align} \label{inv}
& I_1= \text{tr}(\vec{B}), \quad
I_2=\tfrac{1}{2}\left[I_1^2-\text{tr}\left(\vec{B}^2\right)  \right], \quad
I_3=\det(\vec{B})=J^2,  \nonumber \\
& I_4=\vec{M \cdot CM}, \quad
I_5=\vec{M \cdot C}^2\vec{M}, \notag \\
& I_6=\vec{M}'\vec{\cdot CM}', \quad
I_7=\vec{M}'\vec{\cdot C}^2\vec{M}', \quad
I_8=\vec{M \cdot CM}'.
\end{align}

The constraint of incompressibility, requiring that $I_3=1$ for all deformations, is usually imposed for two important classes of materials: elastomers and biological soft tissues. 
In numerical simulations of these materials, however, the material is usually assumed to be almost incompressible instead due to the numerical difficulty in globally enforcing the incompressibility constraint. This also has the virtue of being closer to the physics of these materials, as all materials suffer some volume change when deformed. 

There is no standard formulation of slight compressibility. The usual approach (see, for example, \citealp{Book},  \citealp{Hea}, \citealp{bonet}, and the manuals cited in the References) is first to reformulate the kinematics in terms of the \emph{modified} or \emph{distortional} stretches, $\lambda_i^*$, defined as
\be \label{ns}
\lambda_i^* \equiv J^{-1/3}\lambda_i.
\en
The motivation for doing this is to develop a theory that has close parallels with the now classical theory of perfectly incompressible materials since $\lambda_1^*\lambda_2^*\lambda_3^*=1$. The tensorial measures of deformation can therefore be multiplicatively decomposed into dilatational and volume-preserving parts as follows \citep{Hea}:
$$
\vec{F}=\left(J^{1/3}\vec{I}\right)\vec{F}^*=J^{1/3}\vec{F}^*, \quad \vec{B}=\left(J^{2/3}\vec{I}\right)\vec{B}^*=J^{2/3}\vec{B}^*, \quad \vec{C}=\left(J^{2/3}\vec{I}\right)\vec{C}^*=J^{2/3}\vec{C}^*,
$$
with the relationship between the two sets of invariants $\{I_1, I_2, I_3, I_4, ..., I_8 \}$ and $\{I_1^*, I_2^*, I_4^*,\\ ..., I_8^*;J \}$ given by
\be \label{modi}
I_a^*=J^{-2/3}I_a, \quad a \in \{1,4,6,8\}, \qquad I_b^*=J^{-4/3}I_b, \quad b \in \{2,5,7\}, \qquad I_3^*=1.
\en
In terms of the modified stretches, the stress-strain relation \eqref{cs} becomes
\begin{align} \label{csm}
J\vec{\sigma}= & 2I_3\frac{\partial W}{\partial I_3}\vec{I}+2W_1^*\vec{B}^*+2W_2^*\left[I_1^*\vec{B}^*-\left(\vec{B}^*\right)^2\right]+2W_4^*\vec{F^*M} \otimes \vec{F^*M}  \nonumber \\
 & + 2W_5^*\left( \vec{F^*M} \otimes \vec{B^*F^*M}+\vec{B^*F^*M} \otimes \vec{F^*M} \right) + 2W_6^*\vec{F^*M}^{\prime} \otimes \vec{F^*M}^{\prime}  \nonumber \\
 & + 2W_7^*\left( \vec{F^*M}^{\prime} \otimes \vec{B^*F^*M}^{\prime}+\vec{B^*F^*M}^{\prime} \otimes \vec{F^*M}^{\prime} \right) \nonumber \\
 & + W_8^*\left( \vec{F^*M} \otimes \vec{F^*M}^{\prime}+\vec{F^*M}^{\prime} \otimes \vec{F^*M} \right), 
\end{align}
where, now, the subscripts attached to $W^*=W^*\left(I_1^*, I_2^*, I_4^*,..., I_8^*;J \right)$ denote partial differentiation with respect to the modified invariants $I_c^*$. In terms of the invariants $\{I_1^*, I_2^*, I_4^*,\\ ..., I_8^*;J \}$, the $\partial W/\partial I_3$ term has the form
$$
2I_3\frac{\partial W}{\partial I_3}=J\frac{\partial W^*}{\partial J}-\tfrac{2}{3} W_a^*I_a^*-\tfrac{4}{3} W_b^*I_b^*,
$$
where the ranges of the summation over repeated subscripts are given in \eqref{modi}. Consequently the Cauchy stress-strain relation has the following final form in terms of the normalized invariants and $J$:
\begin{align} \label{final}
J\vec{\sigma} = & J\frac{\partial W^*}{\partial J}\vec{I}+ 2W_1^*\left(\vec{B}^*-\tfrac{1}{3}I_1^*\vec{I}\right)+2W_2^*\left[ I_1^*\vec{B}^*-\left(\vec{B}^*\right)^2-\tfrac{2}{3}I_2^*\vec{I}\right] \nonumber \\
& + 2W_4^*\left(\vec{F^*M} \otimes \vec{F^*M}-\tfrac{1}{3}I_4^*\vec{I}\right) +  \nonumber \\
&+ 2W_5^*\left( \vec{F^*M} \otimes \vec{B^*F^*M}+\vec{B^*F^*M} \otimes \vec{F^*M}-\tfrac{2}{3}I_5^*\vec{I} \right)  \nonumber \\
& + 2W_6^*\left(\vec{F^*M}^{\prime} \otimes \vec{F^*M}^{\prime}-\tfrac{1}{3}I_6^*\vec{I}\right)   \nonumber \\
& + 2W_7^*\left( \vec{F^*M}^{\prime} \otimes \vec{B^*F^*M}^{\prime}+\vec{B^*F^*M}^{\prime} \otimes \vec{F^*M}^{\prime}-\tfrac{2}{3}I_7^*\vec{I} \right)  \nonumber \\
& + W_8^*\left(\vec{F^*M} \otimes \vec{F^*M}^{\prime}+\vec{F^*M}^{\prime} \otimes \vec{F^*M}-\tfrac{2}{3}I_8^*\vec{I}\right). 
\end{align}

Taking the trace of both sides yields the following important property of strain-energies defined in terms of these normalised invariants:
\be \label{tr1}
\text{tr}\left(\vec{\sigma}\right)=3\frac{\partial W^*}{\partial J}.
\en
This term is usually called the \emph{hydrostatic stress}. 
The contrast with the corresponding result for the classical invariants, given by
\be \label{tr2}
J \, \text{tr}\left(\vec{\sigma}\right)=6W_3I_3+2 W_aI_a + 4 W_bI_b,
\en
is immediate. The simplicity of form of the first suggests a more convenient choice of invariants but this very simplicity poses problems for the theory expressed in terms of the normalised invariants as will be shown shortly. 

Although \eqref{final} is the general constitutive form assumed for slightly compressible materials reinforced with two families of fibres, this form is in fact valid for \emph{all} materials reinforced with two families of fibres. The constitutive assumption that is widely assumed to be specific to slightly compressible materials is that the strain-energy function in \eqref{final} can be \emph{additively decomposed} into dilatational and volume preserving parts as:
\be \label{sef}
W^*\left(I_1^*, I_2^*, I_4^* ..., I_8^*;J\right)=f(J) + \mathcal{W}\left(I_1^*, I_2^*, I_4^*, ..., I_8^*\right).
\en
It follows immediately from \eqref{tr1} that these materials have the important property that 
\be \label{hy}
\text{tr}\left(\vec{\sigma}\right)=3f'(J),
\en
and consequently the hydrostatic stress is a function of the $J$ invariant only. It will be seen that this property has implications for the validity of the decomposition \eqref{sef}. The decomposition \eqref{sef} is the usual starting point for Finite Element analysis of anisotropic, non-linearly elastic materials. Most authors (see, for example, \citealp{Hea}, \citealp{Cris}, \citealp{SaT}, \citealp{Wea}) assume this axiomatically without motivation, relying on physical intuition and, almost certainly, mathematical convenience as the basis for this assumption. Its validity will be considered in Section \ref{add}.

%%%%%%%%%%%%%%%%%%%%%%%%%

\section{Isotropic materials}
\label{Isotropic materials}

%%%%%%%%%%%%%%%%%%%%%%%%%

Before considering the anisotropic materials defined by either \eqref{cs} or \eqref{final}, some insight into the validity of assuming \eqref{sef} can be gained by first considering isotropic materials. \cite{San} and \cite{HaM} have shown for isotropic materials that a decomposition of the form 
\be \label{iso}
W^*\left(I_1^*, I_2^*, J\right)=f(J) + \mathcal{W}\left(I_1^*, I_2^*\right),
\en
which is the specialisation of \eqref{sef} to isotropic materials, holds if and only if \eqref{hy} holds \emph{for all deformations}. Condition \eqref{hy} is a generalization to all deformations of the intuitive idea that, for isotropic materials, a pure hydrostatic stress, with the principal stresses all equal, should only cause a volume change and therefore $\text{tr}\left(\vec{\sigma}\right)$ should depend only on $J$.

To represent this intuitive idea mathematically, note that, in terms of the classical Cauchy-Green strain invariants, pure hydrostatic stress for isotropic materials is described by $\vec \sigma = T \vec I$, say, giving
\begin{eqnarray} \label{hys}
JT&=&2W_1\lambda_1^2+2W_2\lambda_1^2\left(\lambda_2^2+\lambda_3^2\right) + 2J^2W_3, \nonumber \\
JT&=&2W_1\lambda_2^2+2W_2\lambda_2^2\left(\lambda_1^2+\lambda_3^2\right) + 2J^2W_3,  \nonumber \\
JT&=&2W_1\lambda_3^2+2W_2\lambda_3^2\left(\lambda_1^2+\lambda_2^2\right) + 2J^2W_3.
\end{eqnarray}
Simple subtraction then yields
$$
0=\left(\lambda_i^2-\lambda_j^2\right)\left(W_1+\lambda_k^2W_2\right).
$$
Assuming that the so-called empirical inequalities,
$W_1>0$, $W_2 \ge0$,
hold means that, for isotropic materials, pure hydrostatic stress and strain are equivalent. For pure hydrostatic stress then, the deformation is a single-parameter deformation and therefore it seems reasonable that pure hydrostatic stress should depend only on one strain invariant and that that invariant should be the volume measuring invariant $J$, i.e., it seems reasonable that \eqref{hy} holds. 

However, even for isotropic materials, requiring that the relation \eqref{hy} holds for all deformations, which is equivalent to the additive decomposition \eqref{iso}, seems overly prescriptive. The mean Cauchy stress should cause more than a simple volume change in general and \eqref{hy} therefore does not seem an \emph{a priori} constitutive assumption rooted in rational mechanics, even though its equivalent \eqref{iso} has some intuitive appeal on the basis of mathematical convenience. 

The modelling of slight compressibility, even for the isotropic case, is severely hindered by the scarcity of relevant experimental data. Although there are classical studies investigating the hydrostatic response of elastomers, most notably those of \cite{AaG} and \cite{Brid}, they involve pressures of the order of hundreds of MPa (see, for example, \citealp{HaM, HaM2}, for recent reviews of the available data on slight compressibility of rubber-like materials) and are thus not relevant when one wishes to consider the effect of slight compressibility under pressures experienced by elastomers in typical applications and by soft tissue under physiological pressure. There is, however, one set of experiments that can be considered the benchmark set of data when it comes to measuring the slight compressibility of elastomers. In a series of elegant experiments on peroxide vulcanizates of natural gum rubber using a dilatometer technique, \cite{Penn} measured the volume change for a number of rubberlike materials in simple tension. His data are summarized and tabulated in \cite{FaP}. Penn's conclusion is that \emph{no model of the form \eqref{iso} is compatible with his data}, although a recent re-analysis of Penn's data \citep{HaM} suggest that a quantitative, if not qualitative, agreement of \eqref{iso} with Penn's data can be achieved. If Penn is correct, however, then the additive decomposition is fatally flawed for isotropic solids.

%%%%%%%%%%%%%%%%%%%%%%%%

\section{Additive decomposition for anisotropic materials} \label{add}

%%%%%%%%%%%%%%%%%%%%%%%%

It is now shown that for the anisotropic materials of interest here, i.e., compressible, non-linear, hyperelastic materials with two preferred directions, the additive decomposition \eqref{sef} is \emph{equivalent} to \eqref{hy}, thus generalising the result for isotropic materials considered in the last section. 

It has already been shown that \eqref{hy} is a necessary consequence of \eqref{sef}. Sufficiency will now be proved. Assume then that \eqref{hy} holds. Then it follows from \eqref{cs} that
\begin{align} \label{trcs}
J\, \text{tr} \left(\vec{\sigma}\right) & =2I_1W_1+4I_2W_2+6I_3W_3+2I_4W_4 + 4I_5W_5 + 2I_6W_6+ 4I_7W_7 + 2I_8 W_8\nonumber \\ &= 3Jf'(J),  
\end{align}
which is a linear partial differential equation in $W$. The particular integral is trivial:
\begin{equation}
W_p=f\left(I_3^{1/2}\right).
\end{equation}
The homogeneous equation has the form
\begin{equation}
I_1W_1+2I_2W_2+3I_3W_3+I_4W_4 + 2I_5W_5 + I_6W_6+ 2I_7W_7 +I_8 W_8=0,
\end{equation}
which has the general solution
\begin{equation}
W_h=\mathcal{W}\left(I_1^*, I_2^*, I_4^*,I_5^*,I_6^*,I_7^*, I_8^*\right),
\end{equation}
where $\mathcal{W}$ is an arbitrary function and the starred invariants are the normalised invariants defined in \eqref{modi}. Consequently the solution of the partial differential equation \eqref{trcs} is given by \eqref{sef} and the equivalence result has been proved.

In summary, if $W=W\left(I_1,I_2,I_3, \dots, I_8\right)$ then 
\begin{equation}
\text{tr}\left(\vec{\sigma}\right)=3f'(J),
\end{equation}
if, and only if,
\begin{equation}
W=f(J) + \mathcal{W}\left(I_1^*, I_2^*, I_4^* ..., I_8^*\right).
\end{equation}
This type of equivalence for anisotropic materials was first obtained by \cite{San}, who considered materials that were orthotropic. 

Some implications of this equivalence are now considered. Although \eqref{sef} might seem a natural assumption to model slight compressibility of anisotropic materials in terms of the strain-energy function, its stress equivalent \eqref{hy} suggests that it is seriously flawed. It was argued previously for isotropic materials that the mean Cauchy stress should cause more than a simple volume change, with the possible exception of pure hydrostatic stress. This argument seems especially valid for anisotropic materials: the effect of the mean Cauchy stress should contain some dependence on the fibre orientation and therefore contain some dependence on the pseudo-invariants $I_4, \dots, I_8$ or their normalised equivalents. In addition, the different responses of the fibres and the matrix should cause some change in shape and consequently some dependence on $I_1,I_2$ must also be needed.  This should be true even for pure hydrostatic stress, for if one considers, say, the pure hydrostatic stress of a cube containing two families of reinforcing fibres, then the fibres will be stiffer than the surrounding matrix. Elementary physics therefore suggests that the cube will be deformed into an hexahedron with non-parallel faces. In fact, for orthotropic materials, \cite{San} proved  that a purely spherical state of stress will be accompanied by a change of shape as well. 
However, our repeated numerical experiments using ABAQUS 6.9 of the hydrostatic tension of slightly compressible cubes show that ABAQUS predicts that the cubic shape is maintained, see Figure \ref{fig:cube}. 

It seems then that the additive decomposition of the strain-energy function into volumetric and deviatoric parts is essentially an \emph{isotropic} condition (and even then appropriate for pure hydrostatic tension only) and was simply extended in a natural way for anisotropic materials without adequate consideration being given to the corresponding physics. The arguments above suggest that, despite its intuitive appeal, it should \emph{not} be employed when modelling nonlinear, anisotropic materials which are characterized by infinitesimal volume changes when deformed. Certainly its equivalent formulation in terms of stresses, \eqref{hy}, does not seem a natural or appropriate constitutive assumption to make when modelling the slight compressiblity of anisotropic materials.   A formulation of the theory that accounts for infinitesimal volume changes in a physically realistic way is badly needed. 

There are some data in the literature that measure the volume change of biological, soft tissue in deformation but they are neither as comprehensive nor as useful as the data of Penn.  \cite{Cea} measured the volume change accompanying the internal pressurization of arteries from dogs and observed volume changes much smaller than $1 \%$. Unfortunately, these data are of limited use for constitutive modelling as the underlying deformation is \emph{in}homogeneous. \cite{CaF} conducted uniaxial compression experiments on rabbit thoracic arteries and also concluded that the tissue is only slightly compressible. However these experiments were one-dimensional and consequently cannot yield information about the interaction of volume change and anisotropy. Consequently extensive, careful experimentation is needed to measure the slight compressibility of soft tissue, with a particular emphasis on the effect of anisotropy. This data must be valid for the entire range of physiological loading and must incorporate as much variety in the type of soft tissue used as possible. It is only when such data are available that reliable models of slight compressibility can be formulated and the reliability and accuracy of Finite Element simulations of the mechanical response of soft tissue be improved. The analysis presented here has illustrated the difficulties of modelling the slight compressibility of soft tissue using intuitive mathematical models.

Even if one is unconvinced by the theoretical arguments against the additive decomposition \eqref{sef} and the claim of \cite{Penn} that his experimental data is incompatible with it, there are serious practical difficulties in simulating slightly compressible anisotropic materials. This is illustrated next using the example of equi-biaxial strain, which locally models the deformation of arterial tissue under internal pressure.

%%%%%%%%%%%%%%%%%%%%

\section{Finite Element simulations}
\label{Finite Element simulations}

%%%%%%%%%%%%%%%%%%%%

We  now consider some Finite Element simulations of the Holzapfel-Gasser-Ogden model of anisotropic, hyperelastic behaviour implemented by ABAQUS (see Section 19.5.3 of the Abaqus Analysis User's Manual). For isothermal deformations, this assumes the strain-energy function can be additively decomposed into a volumetric and deviatoric parts as in \eqref{sef} where, using our previous notation,
\be \label{abv}
f(J)=\frac{1}{D}\left[\tfrac{1}{2}(J^2-1)-\ln J\right],
\en
and 
\begin{multline} \label{abd}
 \mathcal{W}\left(I_1^*, I_2^*, I_4^* ..., I_8^*\right) = C_{10}\left(I_1^*-3\right) \\
 +\frac{k_1}{2k_2}\left\{\text{e}^{k_2\left[\kappa_0 I_1^*+\left(1-3\kappa_0\right)I_4^*-1\right]^2}+\text{e}^{k_2\left[\kappa_0 I_1^*+\left(1-3\kappa_0\right)I_6^*-1\right]^2}-2\right\},
\end{multline}
where $C_{10}$, $k_1$ and $k_2$ are constants and $\kappa_0$ is the dispersion parameter (not to be confused with the `bulk modulus' $\kappa$).

There are two approaches to the specification of the compressibility factor $D$. The first is to treat this factor as a material parameter, with the expectation therefore that there is some experimental justification for its choice. Referring to the ABAQUS manual \cite{ABA}, $D$ is defined as $D=2/\kappa$, where $\kappa$ is the bulk modulus without any definition of what the term `bulk modulus' might mean in an anisotropic context. In fact the manual's definition is circular, $D$ is defined in terms of $\kappa$ and vice versa. What seems implicit in the ABAQUS formulation of compressibility is the existence of a set of experimental data for pure hydrostatic stress, $T$,  for which $T=f'(J)$ in general and for which 
\be \label{Ddef}
T=\frac{1}{D}\left(J-\frac{1}{J}\right),
\en
holds in particular here. Then $D=2/f^{''}(1)$. There are two obvious practical difficulties with this approach: there are no experimental data for biological, soft tissue in hydrostatic stress and even if there were, almost certainly the relation \eqref{Ddef} would not hold as there needs to be some incorporation of the fibre orientation into the hydrostatic response. The important default setting, used in almost all Finite Element simulations in practice, is derived from setting $\kappa/\mu=20$, where $\mu$ is defined to be the largest value of the initial shear moduli. Again, rarely, if ever, are there experimental data available to identify the largest of the shear moduli and, of course, the value of the ratio adopted is arbitrary. It seems to have been imported directly into the modelling of anisotropic materials from the corresponding models of isotropic elasticity, where it correspond to a Poisson's ratio of 0.475, which has been adopted without justification by ABAQUS as being representative of the behaviour of elastomers.

The second approach is to treat $D$ as a penalty parameter in a purely computational approach. A clear exposition of this approach can be found in \cite{Hea1}, where $\kappa$ is a user-specified and mathematically motivated penalty parameter. The authors state that  `an appropriate value for $\kappa$ is determined through numerical experiments', although what an appropriate value might be or what these numerical experiments entail is not discussed. A hint as to the size of the appropriate value is given when the ratio of the penalty parameter used in the simulations to the shear moduli of the modelled material is said to be approximately three orders of magnitude. Interestingly, this is several orders of magnitude greater than the default setting for the first approach used in ABAQUS. However here again $D$ is essentially an arbitrary parameter and it will be shown that varying this parameter in an obvious way has a significant effect on the stress distribution.

The values for the material constants used in our simulations are those proposed by \cite{Gea} to model the mechanical response of the iliac adventitia, i.e.,  
\be \label{conval}
C_{10}=3.82 \text{ kPa}, \quad k_1=996.6 \text{ kPa},  \quad  k_2=524.6, \quad \kappa_0=0.226,
\quad
\Theta=49.98^{\circ},
\en
where $\Theta$ is the angle between the fiber directions $\vec M$ and $\vec M'$ in the reference configuration.
As already discussed, ABAQUS implicitly identifies the anisotropic compressibility factor with its isotropic equivalent, with therefore
$$
D=\frac{3(1-2\nu)}{\mu(1+\nu)},
$$
where $\nu$, $\mu$ are the initial Poisson's ratio and shear modulus, respectively. Taking the shear modulus as $\mu = 2C_{10}= 7.64$ kPa \citep{Gea}, we  use this identification to vary the values of $D$, with units $\left(\text{Pa}\right)^{-1}$, by changing $\nu$ as detailed in Table \ref{Table1}.

\vspace{6cm}

\begin{table}[ht!]
\begin{center} 
\begin{tabular}{|c|c| c| c|} 
\hline
$\nu$ & $D \left(\text{Pa}\right)^{-1}$ & $\kappa \, (\text{Pa})$ & $\kappa/\mu$ \\ 
\hline 
 0.475 & 1.33 $\times 10^{-5}$ & $1.50 \times 10^5$ &20 \\
0.48 &   1.06 $\times 10^{-5}$& $1.88 \times 10^5$ &25 \\
0.49 &    5.27 $\times 10^{-6}$&$3.79 \times 10^5$ &50 \\
0.495 & 2.63$ \times 10^{-6}$&$7.61 \times 10^5$ & 100\\
 0.499 & 5.24 $\times 10^{-7}$& $3.82 \times 10^6$ & 500 \\
 0.4995 &  2.62 $\times 10^{-7}$&$7.64 \times 10^6$ & 1000\\
0.4999 & 5.24 $\times 10^{-8}$&$3.82 \times 10^7$ & 5000 \\
\hline 
\end{tabular} 
\caption{\label{Table1} The different degrees of compressibility for arteries under equi-biaxial strain.
See Figs. \ref{T11} and \ref{T22} for the corresponding Cauchy-stress response.}
\end{center} 
\end{table}

A square sample of dimensions 10 mm $\times $ 10 mm and thickness 0.5 mm was modelled using 8320 reduced integration hexahedral (C3D8R) elements for the mesh. The numerical analyses were performed using both the static analysis procedure in ABAQUS/Stan\-dard 6.9 and the dynamic, explicit procedure in ABAQUS/Explicit 6.9, with both procedures giving identical results for slight compressibility. The perfectly incompressible case was modelled using ABAQUS/Standard only, since it is not possible to model the incompressible case using ABAQUS/Explicit. Displacement was applied through a constant velocity  boundary condition applied to two sides of the square while the other two sides were fully constrained in all six degrees of freedom, resulting in an equibiaxial deformation. The simulation results were independent of mesh density, element type and sample thickness.

\begin{figure} [htp!]
\begin{center}
\includegraphics*[width=17cm]{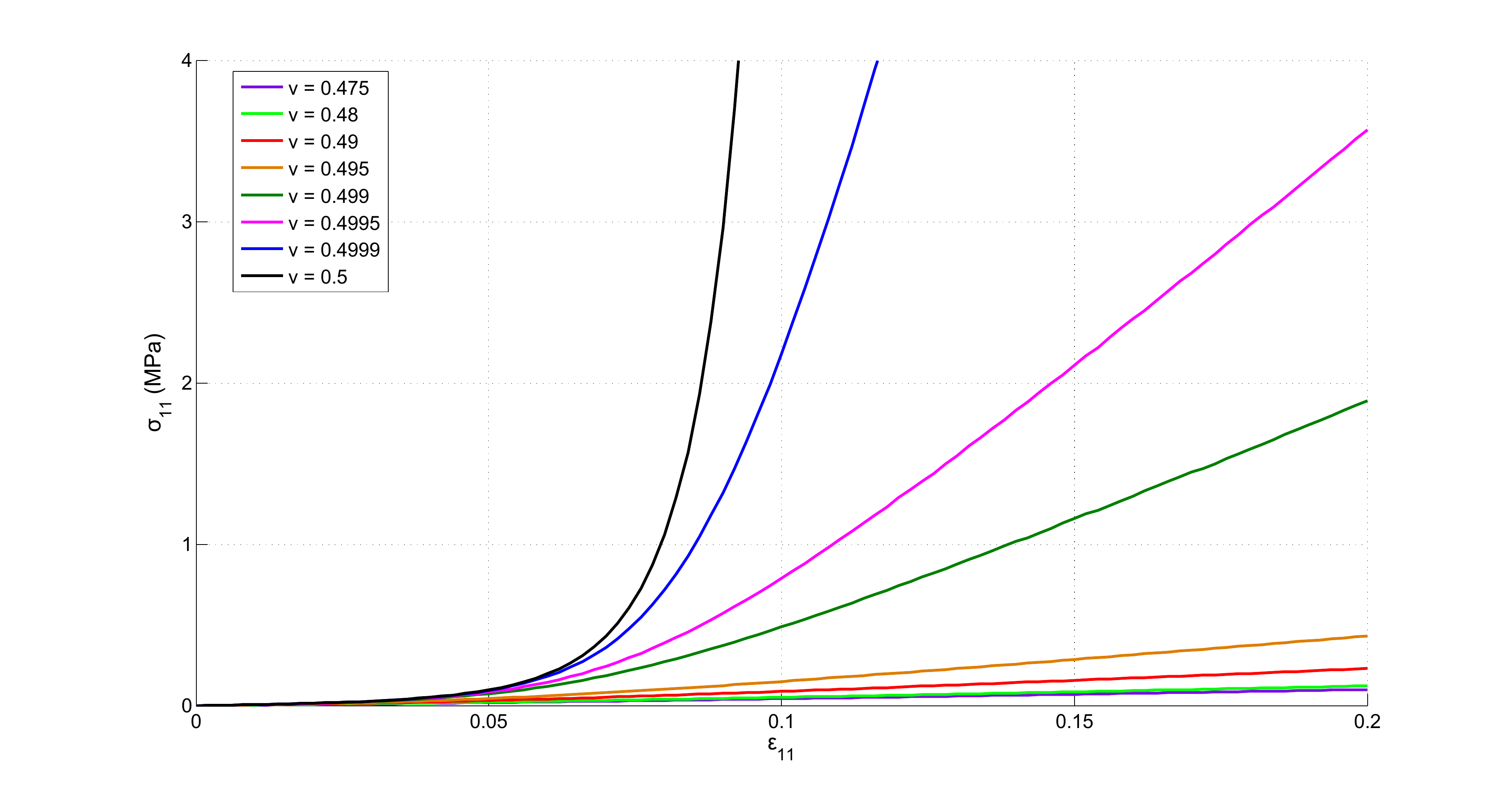}
\end{center}
\caption{Variation in Cauchy stress in 1-direction for different values of Poisson's ratio.}
\label{T11}
\end{figure}

\begin{figure} [htp!]
\begin{center}
\includegraphics*[width=17cm]{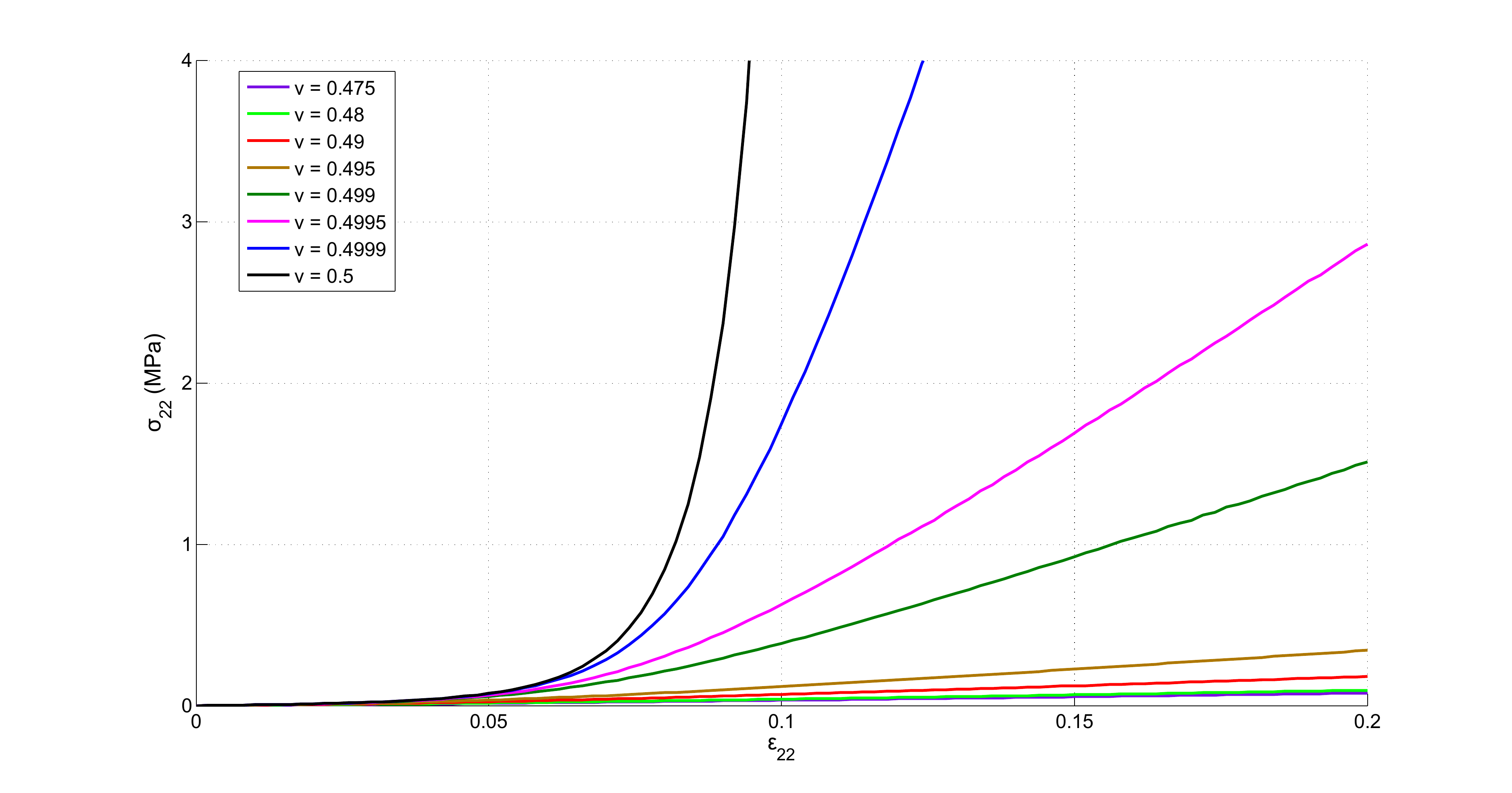}
\end{center}
\caption{Variation in Cauchy stress in 2-direction for different values of Poisson's ratio.}
\label{T22}
\end{figure}

Fig.~\ref{T11} illustrates the variation in Cauchy stress in the 1-direction throughout the duration of the simulation. Similarly, Fig.~\ref{T22} illustrates the variation in Cauchy stress in the 2-direction, which, despite the equibiaxial nature of the simulation, differs from Fig.~\ref{T11} because of the anisotropic nature of the material. It can be seen that the Cauchy stress varies dramatically up to a strain of 20\% depending on the chosen value of Poisson's ratio.

The implications of these plots for the FE analysis of biological soft tissue are obvious: there is a varation of several orders of magnitude in the predicted stresses when Poisson's ratio is varied between $0.475$, the default value in ABAQUS, and $0.5$, the incompressible limit, even for strains as low as $5-10 \%$, which are typical physiologically. The correct value (remember we are assuming that the additive strain decomposition is valid) of Poisson's ratio is never known in practice and there will almost certainly be a difference between the assumed and actual values, resulting in a significant error in the predicted stresses. The seemingly obvious solution to this extreme sensitivity to Poisson's ratio of simply simulating perfect incompressibility is not a valid approach. No material is perfectly incompressible and assuming this idealisation (denoted in our results by $\nu=1/2$) will result in a difference between the idealised value assumed and the actual value of Poisson's ratio.

There is another feature of these plots worth noting: the exponential form of the strain-energy function \eqref{abd} was originally proposed by \cite{Hea2} as a means of modelling the severe anisotropic strain-hardening observed in experiments on biological, soft tissue. The above plots suggest that this strain-hardening feature of the model is blunted with decreasing values of Poisson's ratio. Indeed for $\nu = 0.475$, a not untypical value for elastomers \citep{BaS}, the strain-hardening effect is moderate at best. Thus if one were to assume a compressibility parameter close to this value of Poisson's ratio, the essential severe strain-hardening effect will not be reflected in the simulations. 

%%%%%%%%%%%%%%%%%%%%%%%%%

\section{Stress sensitivity} \label{sen}

%%%%%%%%%%%%%%%%%%%%%%%%%

The source of this hyper-sensitivity to variation in the value of Poisson's ratio is now explored.  Consider first the response of slightly compressible, anisotropic materials to general pure homogeneous deformations of the form
$$
x=\lambda_1X, \quad y = \lambda_2Y, \quad z=\lambda_3Z,
$$
where $\left(X,Y,Z\right)$ and $\left(x,y,z\right)$ are the Cartesian coordinates of a typical particle before and after deformation, respectively, and the $\lambda$s are positive constants. For simplicity, it is further assumed that the directions of anisotropy in the undeformed configuration are planar and symmetric with respect to each other of the form
$$
\vec{M}=C\vec{I} + S\vec{J}, \quad \vec{M}'=C\vec{I} - S\vec{J}, \quad \text{where } C \equiv \cos \Phi, \quad S \equiv \sin \Phi, \quad 0<\Theta<\pi/2,
$$
using an obvious notation for the unit vectors in the undeformed configuration. The fibre directions are deformed into the vectors $\vec{FM}, \, \vec{FM}'$ given by
$$
\vec{FM}=\lambda_1C\vec{i} + \lambda_2S\vec{j}, \quad \vec{FM}'=\lambda_1C\vec{i} - \lambda_2S\vec{j}.
$$
In terms of the normalized stretches \eqref{ns} and their corresponding invariants \eqref{modi}, the principal Cauchy stresses for pure homogeneous deformations for two families of plane mechanically equivalent fibres can be obtained from \eqref{final} and are given by
\begin{align} \label{csn}
J\sigma_{1}= &J\frac{\partial W^*}{\partial J}  + \tfrac{2}{3}W_1^*\left( 2\lambda_1^{*2}-\lambda_2^{*2}-\lambda_3^{*2}\right) + \tfrac{2}{3}W_2^*\left( \lambda_1^{*2}\lambda_2^{*2}+ \lambda_1^{*2}\lambda_3^{*2}-2 \lambda_2^{*2}\lambda_3^{*2}\right)  \nonumber \\
&+ \tfrac{2}{3}\left(W_4^*+W_6^*\right)\left(2\lambda_1^{*2}C^2-\lambda_2^{*2}S^2\right) + \tfrac{4}{3}\left(W_5^*+W_7^*\right)\left(2\lambda_1^{*4}C^2-\lambda_2^{*4}S^2\right)  \nonumber \\
& + \tfrac{2}{3}W_8^*\left(2\lambda_1^{*2}C^2+\lambda_2^{*2}S^2\right), \nonumber \\[4pt]
J\sigma_{2} = &J\frac{\partial W^*}{\partial J}  + \tfrac{2}{3}W_1^*\left( 2\lambda_2^{*2}-\lambda_1^{*2}-\lambda_3^{*2}\right) + \tfrac{2}{3}W_2^*\left( \lambda_1^{*2}\lambda_2^{*2}+ \lambda_2^{*2}\lambda_3^{*2}-2 \lambda_1^{*2}\lambda_3^{*2}\right)  \nonumber \\
& + \tfrac{2}{3}\left(W_4^*+W_6^*\right)\left(2\lambda_2^{*2}S^2-\lambda_1^{*2}C^2\right) + \tfrac{4}{3}\left(W_5^*+W_7^*\right)\left(2\lambda_2^{*4}S^2-\lambda_1^{*4}C^2\right) \nonumber \\
& - \tfrac{2}{3}W_8^*\left(2\lambda_2^{*2}S^2+\lambda_1^{*2}C^2\right), \nonumber \\[4pt]
J\sigma_{3}=&J\frac{\partial W^*}{\partial J}  + \frac{2}{3}W_1^*\left( 2\lambda_3^{*2}-\lambda_1^{*2}-\lambda_2^{*2}\right) + \tfrac{2}{3}W_2^*\left( \lambda_1^{*2}\lambda_3^{*2}+ \lambda_2^{*2}\lambda_3^{*2}-2 \lambda_1^{*2}\lambda_2^{*2}\right)  \nonumber \\
& -\tfrac{2}{3}\left(W_4^*+W_6^*\right)\left(\lambda_1^{*2}C^2+\lambda_2^{*2}S^2\right) - \tfrac{4}{3}\left(W_5^*+W_7^*\right)\left(\lambda_1^{*4}C^2+\lambda_2^{*4}S^2\right)  \nonumber \\
&+\tfrac{2}{3}W_8^*\left(\lambda_2^{*2}S^2-\lambda_1^{*2}C^2\right),
\end{align}
which for plane-stress experiments become
\begin{eqnarray} \label{pse}
J\sigma_{1}&=& 2W_1^*\left( \lambda_1^{*2}-\frac{1}{\lambda_1^{*2}\lambda_2^{*2}}\right) + 2\lambda_2^{*2}W_2^*\left( \lambda_1^{*2}-\frac{1}{\lambda_1^{*2}\lambda_2^{*2}}\right) + 2\left(W_4^*+W_6^*+W_8^*\right)\lambda_1^{*2}C^2 \nonumber \\
&& + 4\left(W_5^*+W_7^*\right)\lambda_1^{*4}C^2, \nonumber \\
J\sigma_{2}&=& 2W_1^*\left( \lambda_2^{*2}-\frac{1}{\lambda_1^{*2}\lambda_2^{*2}}\right) + 2\lambda_1^{*2}W_2^*\left( \lambda_2^{*2}-\frac{1}{\lambda_1^{*2}\lambda_2^{*2}}\right) + 2\left(W_4^*+W_6^*+W_8^*\right)\lambda_2^{*2}S^2 \nonumber \\
&& + 4\left(W_5^*+W_7^*\right)\lambda_2^{*4}S^2,
\end{eqnarray}
noting that $\lambda_3^*=1/\left(\lambda_1^*\lambda_2^*\right)$. For equi-biaxial \emph{strain}, plane-stress experiments, these principal stresses therefore become
\begin{eqnarray} \label{p1}
J\sigma_{1}&=& 2W_1^*\left( \lambda_1^{*2}-\frac{1}{\lambda_1^{*4}}\right) + 2\lambda_1^{*2}W_2^*\left( \lambda_1^{*2}-\frac{1}{\lambda_1^{*4}}\right) + 2\left(W_4^*+W_6^*+W_8^*\right)\lambda_1^{*2}C^2 \nonumber \\
&& + 4\left(W_5^*+W_7^*\right)\lambda_1^{*4}C^2, \nonumber \\
J\sigma_{2}&=& 2W_1^*\left( \lambda_1^{*2}-\frac{1}{\lambda_1^{*4}}\right) + 2\lambda_1^{*2}W_2^*\left( \lambda_1^{*2}-\frac{1}{\lambda_1^{*4}}\right) + 2\left(W_4^*+W_6^*+W_8^*\right)\lambda_1^{*2}S^2 \nonumber \\
&& + 4\left(W_5^*+W_7^*\right)\lambda_1^{*4}S^2.
\end{eqnarray}
The corresponding principal stresses for materials assumed to be \emph{perfectly incompressible} are given by
\begin{align} \label{ebs}
& \sigma_1=2W_1\left(\lambda_1^2-\lambda_1^{-4}\right)+2W_2\left(\lambda_1^4-\lambda_1^{-2}\right) + 2\left(W_4+W_6+W_8\right)\lambda_1^2C^2 +  4\left(W_5+W_7\right)\lambda_1^4C^2,   \nonumber \\
& \sigma_2=2W_1\left(\lambda_1^2-\lambda_1^{-4}\right)+2W_2\left(\lambda_1^4-\lambda_1^{-2}\right)+ 2\left(W_4+W_6-W_8\right)\lambda_1^2S^2 +  4\left(W_5+W_7\right)\lambda_1^4S^2.   \nonumber \\
\end{align}
A comparison of \eqref{p1} and \eqref{ebs} reveals the same structure, with an interchange of stretches and normalized stretches resulting in an interchange in the two formulations. This is by design and is an attractive feature of the additive decomposition \eqref{sef}.

Strain-energy functions of the form \eqref{abd}, \eqref{conval} were simulated in the previous section and are now considered here. 
To facilitate the analysis, which is easily generalized in a natural way, it is assumed that 
\be \label{happy}
W^*=\frac{1}{D}\left[\tfrac{1}{2}(J^2-1)-\ln J\right] + \mathcal{W}\left(I_1^*,I_4^*,I_6^*\right),
\en
a general form of the strain-energy function previously considered, and that $\Theta=45^{\circ}$. The latter assumption is a close approximation to the simulated value in \eqref{conval} and simplifies the analysis because now the two principal stress are the same and for slightly compressible materials are given by
\begin{eqnarray} \label{prs}
J\sigma_1 = J \sigma_2= 2\mathcal{W}_1\left( \lambda_1^{*2}-\frac{1}{\lambda_1^{*4}}\right) + \left(\mathcal{W}_4+\mathcal{W}_6\right)\lambda_1^{*2}.
\end{eqnarray}
The corresponding principal stress for perfectly incompressible materials for reduced strain-energy functions of the form $W=W\left(I_1,I_4,I_6\right)$ has the form
\begin{eqnarray} \label{p10}
\sigma_1 = \sigma_2 = 2W_1\left( \lambda_1^{2}-{\lambda_1^{-4}}\right) + \left(W_4+W_6\right)\lambda_1^{2}.
\end{eqnarray}
For strain-energy functions of the form \eqref{happy}, the plane stress condition for equi-biaxial strain with $\Theta=45^{\circ}$ follows from \eqref{csn}$_3$ and is given by
\begin{eqnarray} \label{pss}
0&=&3\left(J^2-1\right)  +2D\left[2\mathcal{W}_1\left( \frac{1}{\lambda_1^{*4}}-\lambda_1^{*2}\right) - \left(\mathcal{W}_4+\mathcal{W}_6\right)\lambda_1^{*2}\right].
\end{eqnarray}
As before, let $\mu$  be the largest value of the initial shear moduli. Let $\tilde{D} \equiv D\mu$, a non-dimensional measure of the relative contribution of the volumetric and deviatoric parts of the strain-energy function. Then 
\begin{eqnarray} \label{dm}
0&=&3\left(J^2-1\right)  +2\tilde{D}\left[2\hat{\mathcal{W}}_1\left( \frac{1}{\lambda_1^{*4}}-\lambda_1^{*2}\right) - \left(\hat{\mathcal{W}}_4+\hat{\mathcal{W}}_6\right)\lambda_1^{*2}\right], \quad \hat{\mathcal{W}}_a \equiv \mathcal{W}_a/\mu.
\end{eqnarray}
Given that a typical shear modulus of biological, soft tissue is of the order of kPa, it can be seen from Table 1 that the product  $\tilde{D} = D\mu$ is almost certainly infinitesimal. Motivated by this and the form of the leading-order term in \eqref{dm}, assume then that
\be \label{ja}
J=1+\tilde{D}J^{(1)}=1+\epsilon, \quad \epsilon \ll 1,
\en
i.e., the volume change in infinitesimal. Then 
\be \label{ee1}
\lambda_1^{*}=J^{-1/3}\lambda_1=\left(1+\epsilon\right)^{-1/3}\lambda_1=\lambda_1\left(1-\frac{\epsilon}{3}\right),
\en
neglecting higher order terms here and elsewhere, and 
\be \label{ee2}
I_a^*=J^{-2/3}I_a =I_a\left(1-\frac{2\epsilon}{3}\right), \quad a = 1,4,6.
\en
The leading order term in \eqref{pss} is then of $\mathcal O (\tilde{D})$ and has the form
\be \label{volc}
3J^{(1)}=\sigma_i/\mu,
\en
where $\sigma_i$ is the principal stress for  perfectly incompressible materials, assuming 
$$
\mathcal{W}_a\left(I_1,I_4,I_6\right)=W_a\left(I_1,I_4,I_6\right), \quad a=1,4,6.
$$
 Thus the predicted volume change is simply a third of the corresponding normalised incompressible stress, an interesting consequence of the additive decomposition \eqref{sef}. Note that the  predicted volume change is positive, in agreement with the only relevant experimental data available in the literature obtained by \cite{Penn} and  \cite{CaH} for simple tension experiments on elastomers. With this determination of the volume change, the smallness parameter $\epsilon$ can be interpreted as
$$
\epsilon=\frac{D\sigma_i}{3}.
$$

Return now to the form for the  principal stress for slightly compressible materials given by \eqref{prs}. Expanding the right-hand side in terms of $\epsilon$ and neglecting higher-order terms yields
\be \label{ee}
\sigma_s=\sigma_i-\frac{\epsilon}{3}\,\mathcal{T}
\en
where
\begin{multline}
\mathcal{T} = 2W_1\left(5\lambda_1^2+\frac{1}{\lambda_1^4}\right)+4\left(\lambda_1^2-\frac{1}{\lambda_1^4}\right)\left(I_1W_{11}+I_4W_{14}+I_6W_{16}\right) \nonumber \\
+\lambda_1^2\left[5W_4+5W_6+2I_1W_{14}+2I_4W_{44}+2\left(I_4+I_6\right)W_{46}+2I_1W_{16}+2I_6W_{66}\right].
\end{multline}
A reassuring feature of this analysis is that for $\lambda_1>1$ and for positive partial derivatives of $W$, which is the rule in practice, $\mathcal{T}>0$. Thus the analysis predicts that the  stress will be smaller in the slightly compressible case than in the perfectly incompressible case, which is reflected in the plots of the last section. The truncated Maclaurin series \eqref{ee} has the alternative form
\be \label{df}
\sigma_s=\sigma_i\left(1-\frac{D}{9}\,\mathcal{T}\right).
\en
The hyper-sensitivity of the  stress for slightly compressible materials to variations in Poisson's ratio is thus explained. The data in Table 1 can be summarized as noting that a reduction of one in the second decimal place in Poisson's ratio increases $D$ by an order of magnitude. It follows from \eqref{df} that an increase in $D$ causes a \emph{decrease} in the stress and that this decrease will be significant due to the large change in the value of $D$. This is exactly what is illustrated in Figures \ref{T11}, \ref{T22}.  Note also that the perturbation about the  stress obtained in the perfectly incompressible case is a multiplicative factor rather than the usual additive model. This is a consequence of the decomposition \eqref{sef} and will tend to magnify the variation from the perfectly incompressible case due to variations in the choice of $\nu$.

%%%%%%%%%%%%%%%%%%%%%%%%%

\section{Conclusion}

%%%%%%%%%%%%%%%%%%%

Two problems with the current approach of simulating slightly compressible anisotropic hyperelasticity have been identified. The first is what we believe is a fundamental flaw in the constitutive modelling. The usual decomposition of the strain-energy function into separate volumetric and deviatoric parts is equivalent to assuming that the hydrostatic stress is a function only of the invariant that measures volume change. Thus if one were to consider pure hydrostatic stress of a cube, say, of such a material, the cubic shape would be maintained. This is at variance with simple physics, where, under hydrostatic stress, the fibres and matrix would deform differently, resulting in a change in the cubic shape. Some indication of the presence of the fibres must be reflected therefore in the constitutive model for hydrostatic stress.  

A second problem is a practical problem with the current formulation of slightly compressible, anisotropic materials in commercial finite element codes. The compressibility factor is treated as if it were the reciprocal of the isotropic, infinitesimal bulk modulus. Small variations in the choice of Poisson's ratio close to the limiting value of $1/2$ result in significant variations in the predicted stress. For relatively low levels of strain  ($<$ 10\%) such as commonly occur in soft biological tissue, these predicted stresses can be in error by more than a single order of magnitude. This means that simulated stress distributions based on an arbitrary choice of Poisson's ratio `close' to $1/2$ are very likely to be unrealistic. Whether the predicted stresses are significantly greater than or less than the actual stresses has real implications since a medical device designed using FE tools could be either over designed or, conversely, likely to fail in service. Furthermore, the present work has  demonstrated that the severe strain-hardening effect that seems an essential part of constitutive models of biological, soft tissue is not reflected in models using a Poisson's ratio as high as $0.4995$. Extensive and careful experimentation on the compressibility of biological, soft tissue is needed before robust and reliable models of slight compressibility can be formulated. These will then lead to physically accurate computer based design models for future medical devices that interact with or are embedded into soft biological tissue.

%%%%%%%%%%%%%%%%%%%%%%%%

\section{Acknowledgements}

%%%%%%%%%%%%%%%%%%%%%%%

The authors would like to thank the anonymous referees for their constructive comments. Their insights have been incorporated into this final version of the paper.

%%%%%%%%%%%%%%%%%%%%%

\bibliographystyle{harv}

\end{document}